\documentclass{article}

\usepackage{enumitem}
\usepackage{multirow}
\usepackage{color,todonotes}

\usepackage{microtype}
\usepackage{graphicx}
\usepackage{subfigure}
\usepackage{booktabs} % for professional tables

\usepackage{hyperref}

\usepackage[accepted]{icml2021}

\icmltitlerunning{Predicting students’ performance in online courses using multiple data sources}

\begin{document}

\twocolumn[
\icmltitle{Predicting students’ performance in online courses using multiple data sources}




\begin{icmlauthorlist}
\icmlauthor{Mélina Verger}{upsaclay}
\icmlauthor{Hugo Jair Escalante}{inaoe}
\end{icmlauthorlist}

\icmlaffiliation{upsaclay}{Paris-Saclay University, Orsay, France.}
\icmlaffiliation{inaoe}{Instituto Nacional de Astrofísica, Óptica y Electrónica, Puebla, Mexico}

\icmlcorrespondingauthor{Mélina Verger}{meverger@gmail.com}
\icmlcorrespondingauthor{Hugo Jair Escalante}{hugojair@inaoep.mx}
% melina.verger@universite-paris-saclay.fr

\icmlkeywords{Machine Learning, ICML}

\vskip 0.3in
]



\printAffiliationsAndNotice{} % otherwise use the standard text.

\begin{abstract}
Data-driven decision making is serving and transforming education. We approached the problem of predicting students' performance by using multiple data sources which came from online courses, including one we created.
%Regarding the students’ performance prediction, we approached the problem by using multiple data sources which came from online courses, including one we created. 
Experimental results show preliminary conclusions towards which data are to be considered for the task.
\end{abstract}

\section{Introduction}

Predicting students’ performance is a multidisciplinary task which has a significant impact to many stakeholders, including students, teachers, and educational institutions \cite{Alamri2021}. Not only it helps to predict at-risk students to plan interventions for academic support and guidance \cite{Kotsiantis2004}, but it also helps to identify key factors affecting students’ performance \cite{Alamri2021}. It has then become one of the most challenging issues for educational institutions \cite{Rastrollo-Guerrero2020} regarding the improvement of the teaching-learning process and the more accurate allocation of resources and instructors \cite{Hellas2018}.

During the last decade, such a predictive task has gained more and more attention since online courses along with e-learning resources and educational software had created a new context known as web-based education \cite{Romero2010} in which learners generate an increasing volume and variety of types of data \cite{Papadogiannis2020, Romero2020}.

%Thus, 
Therefore, the problem has been approached by considering a variety of data sources, including, academic, behavioral, demographic and personality, where these sources have been used separately and combined. The benefits of predicting students’ performance using multiple data sources are twofold. On  the one hand, using several data sources enables to discover relationships among different factors and which of these are associated with students’ performance, worthwhile at both education and prediction levels. On the other hand, it fills the gap towards the consideration of a more holistic view of student combining non-cognitive and cognitive characteristics of students that few studies address \cite{Hellas2018}. Most of the articles reviewed by \cite{Hellas2018} have used academic data for prediction. Behavioral data are also common predictor \cite{Alamri2021}. However, \cite{Villagra-Arnedo2016} states that the use of heterogeneous data enriches the final performance of the prediction algorithms. %\hj{Accordingly, we approach the problem by considering multiple information sources. } \todo[author=HJ]{Include something like this...}

Despite the increasing volume of educational data, the Educational Data Mining (EDM) field is still a growing research area where the community has pointed out the scarceness of open data as one of its biggest challenges. In most cases, data comes from the research group conducting the studies \cite{Vijayalakshmi2019} \cite{Baker2009} and is rarely made publicly available for preserving privacy of involved students and teachers. Moreover, most experiments concern a specific context, formed by a population from a single institution and even from a single course \cite{Hellas2018}, which is detrimental to generalizing conclusions with quality confidence because of presumable latent factors \cite{Romero2007, Romero2010}.

From these remarks, this work explores a generic approach to address students’ performance prediction with the aim of answering the following research questions: 
\begin{enumerate}[noitemsep, topsep=0pt]
\item To what extent using multiple data sources on learners increases prediction performance?
\item Which learners’ characteristics are most related to higher academic performance?
\end{enumerate}

In trying to answer these questions, we have performed a study on predicting students' performance as follows. We created an open learning framework to collect learners' data from multiple sources not offered in current open datasets. We studied the collected data with other open datasets focusing on the common features involved. We compared prediction performances using usual data sources but also using the additional ones we collected. We complemented the previous prediction results with feature selection analyses.
%To that purpose, the global organization of our experiments is the following: 
%We created an open learning framework to collect data from multiple sources not offered in current open data, we studied the collected data with other open datasets focusing on the common features involved, we compared prediction performances using usual data sources but also using the additional ones we collected, and we finally complemented the previous prediction results with feature selection analyses.

The global results show that we obtained encouraging similar trends and knowledge with different students’ populations and courses. We highlight the relevance and importance of academic and behavioral data in contrast to demographic ones. The optimal number of features appears to be 3 and we formulated an hypothetical advantageous relationship between students’ who have stronger abilities of memorizing and deducing knowledge from written material and better performance. Additional experiments involving larger datasets are required in order to reach conclusive findings. However, this research work paves the way for the study of new correlations between students' characteristics and their performance.

The remainder of this report is organized as follows. Section II describes the literature review; Section III presents the methodology and the experiments framework; Section IV discusses the results; and Section V provides concluding remarks.

\section{Related work}

\subsection{Students’ performance measures}

Students’ performance could be defined with different measures: exam grade or range, course grade or range, GPA, retention or dropout rate, knowledge gain \cite{Hellas2018}. The choice of a particular metric depends on the educational framework (e.g., in university or in private institution, online or in-person teaching, etc.) and for what it is being predicted. The course grade range (e.g., pass/fail, high-medium-low performance, A-B-C-D-E-F grade) corresponds to the most generic case where instructors are interested in predicting which students will rather belong to a certain class than obtain a precise grade. It is more suitable for adapting their guidance to smaller groups of students. In this work, we focus on pass/fail course grade range which is common with other datasets from online courses framework to make some comparisons.

\subsection{Data sources categories}

\cite{Hellas2018} performed a systematic review based on 357 articles and identified five data sources categories: demographic, personality, academic, behavioral, and institutional. Demographic data includes age, gender, academic level, location, health information, family background, etc. Personality data encompasses psychological data (e.g., self-regulation, self-efficacy, big five personality traits \cite{Goldberg1990}) and affective data (e.g., happiness, anxiety, depression). Academic data covers course performance and/or past performance, during high school for example. Behavioral data includes time on task, number of interactions with the learning material, etc. Institutional data are about teaching approach, high-school quality, etc

Despite this diversity of data, most of the related studies have used either one data source, academic \cite{Hellas2018} or behavioral \cite{Alamri2021}, or at most two data sources with the combination of demographic and academic data. \cite{Villagra-Arnedo2016} highlights an availability-relevance trade-off of educational data which can limit considering more data sources for prediction tasks in the field. Nonetheless, \cite{Villagra-Arnedo2016} states that the use of heterogeneous significant data enriches the final performance of the prediction algorithms, deduced from their experiments with either behavioral features, academic features, both of them or the most correlated ones. \cite{Kotsiantis2004} supports this idea since their highest accuracies were obtained with six machine learning algorithms by adding features progressively. Therefore, this work is intended to be a comprehensive study of all the possible data sources except the institutional one as it is a cross-population experiment without a specific institution. More details of the features involved are provided in subsection \ref{features}. Moreover, considering several data sources enables to discover relationships among different factors and which of these are associated with students’ performance, with a more holistic view of students \cite{Hellas2018}.

\subsection{Data sources availability}

Most of experiments in EDM are conducted with data provided by a university associated to the research group, such as \cite{Mesaric2016, Osmanbegovic2012, Al-Barrak2016, Al-Shehri2017}. The access authorization process as well as privacy preserving mechanisms make difficult to publish the data, which is still a current concern discussed in the latest 
International Conference on Educational Data Mining\footnote{\url{https://educationaldatamining.org/edm2021/}}. Taking this into account with the lack of description on features used \cite{Hellas2018}, it is challenging to select relevant methodology, models, and results among every one of those that are proposed. To make our approach transparent, we have created a learning framework open to anyone and where the anonymized data are publicly available with the consent of the learners (see \url{https://github.com/melinaverger/ed_project}).

\subsection{Explainability purpose}

The very active research area on explainable AI (XAI) and in particular some works focusing on predicting students’ performance, call to use explainable models to equally consider the correctness of the predicted value and the understandability of the prediction made \cite{Alamri2021}, contributing to generalize conclusions in spite of limited data availability. In addition, the recent release of the General Data Protection Regulation (GDPR) re-emphasized the importance of explainable and trustworthy AI when the decision could affect individuals \cite{LongoL.GoebelR.LecueF.KiesebergP.2020}. Therefore, among the most used learning algorithms in EDM (e.g., decision tree, logistic regression, naïve bayes, random forest, support vector machine, k-nearest neighbor, feed-forward neural network, deep neural network \cite{Romero2010, Hellas2018}), we describe benchmark results with explainable models in the next paragraph.

For a binary classification (e.g., pass or fail target), \cite{Kotsiantis2004} used combinations of demographic and academic data with a C4.5 decision tree and reported accuracies of 79.22\% and 73.99\% over 354 and 28 instances respectively, the latter representing a real-condition class. The average over all feature combinations for the 28-instance experiment is 63.43\% accuracy, 69.79\% sensitivity and 57.67\% specificity.  For a 3-class classification (e.g., high, medium or low performance targets), \cite{Vijayalakshmi2019} used demographic, academic and behavioral features and reported an accuracy of 69\% using a C5.0 decision tree on a dataset with 480 instances. For a 4-class classification, \cite{Lakshmi2013} used demographic and academic features and compared ID3, C4.5 and CART decision trees algorithms over 120 instances. C4.5 achieved the highest accuracy at 55.83\%. Similarly, \cite{Chaudhury2016} obtained an average over the four classes of 83.25\% F-score with C4.5 over 648 instances. In this work, we will continue to use an explainable approach and put our results into perspective for the related combinations.

Finally, this work attempts to address some limitations of current state of the art, such as scarceness of open data, single or few data sources, single-context experiment and non-explainable methods and results, by proposing an approach described in the following section.

\section{Methods}
\label{methods}

Summarizing this work, we created an open learning framework to collect data from multiple sources, and we performed the following: (1) we positioned the collected data with two open datasets regarding the common features involved, (2) we compared prediction performances using usual data sources (3) but also using the additional ones we collected, and (4) we  analyzed the previous prediction performances with a feature selection study. This section  provides information on how we collected data, how we extracted features and which algorithms we used for the binary classification prediction task and for the feature selection analyses.

\subsection{Data collection}
\label{data_collection}

\begin{figure}[!ht]
\vskip 0.1in
\begin{center}
\centerline{\includegraphics[width=\columnwidth]{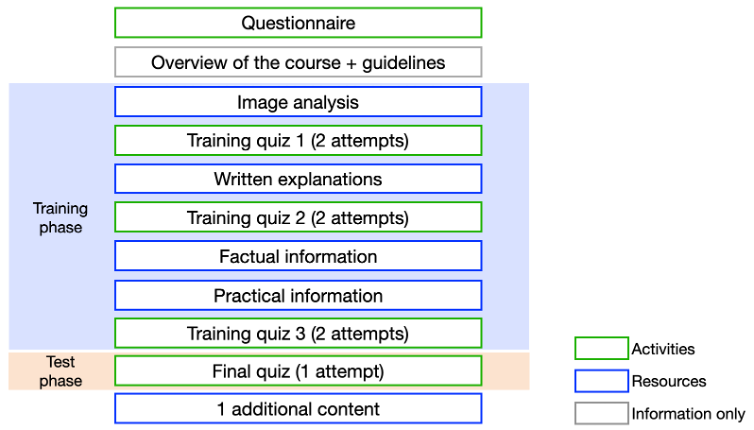}}
\caption{Structure of the course. Uniformly with e-learning platforms common vocabulary, we call \textit{activity} any content where participants need to interact with the course (questionnaires, quizzes, etc.) and \textit{resource} any content where participants only need to consult it (mostly learning material).}
\label{fig-structure}
\end{center}
\vskip -0.2in
\end{figure}

\begin{figure}[!ht]
\vskip 0.1in
\begin{center}
\centerline{\includegraphics[width=\columnwidth]{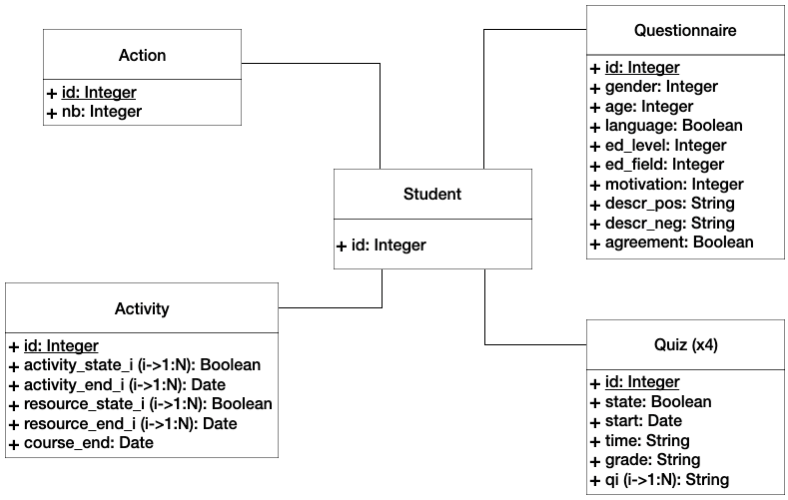}}
\caption{Relational model of anonymized and converted data (73 variables).}
\label{fig-relational_data}
\end{center}
\vskip -0.2in
\end{figure}

As the goal of this work is to exploit multiple data sources not offered in current open data sets, we created an open course on the e-learning platform The Open University\footnote{\url{https://www.open.ac.uk}} (OU) to collect usual and additional features described in subsection \ref{features}. This 1-hour general course is an introduction to the organization of the brain, its characteristics, and functionalities, which does not require substantial prior knowledge. It is thus intended for a general audience with any education background. Each participant had been previously informed of the data collection and the use of the data and had explicitly consented to participate in the proposed learning experiment. This complies with the OU privacy policy\footnote{\url{https://www.open.ac.uk/about/main/strategy-and-policies/policies-and-statements/website-privacy-ou}} and GDPR directives\footnote{\url{https://gdpr-info.eu}}.

The structure of the course is presented in Figure \ref{fig-structure}. It was designed as follows. The duration of the course is a trade-off between involving participants in a learning process with enough time and material, capturing their learning process in equal conditions to others with limited duration and demanding volunteer participants’ efforts with reasonable duration. In Figure 1, it is observed that each resource concerns either a particular format type or content type of the learning material followed right after by a corresponding quiz. Then, each quiz is composed by half of the questions on memorizing information and the other half on deducing knowledge. It must be noted that academic data in this learning framework, e.g., the training quiz grades, are coherent with the final grade, also derived from a quiz, but consequently specific to this final evaluation material.

Besides, since the OU offers a Moodle-based learning management system (LMS) \cite{Watson2007AnAF}, the course permits the collection of the raw data whose variables are described in Figure \ref{fig-relational_data}, after being anonymized and converted in appropriate formats.

\subsection{Feature preprocessing}
\label{features}

\subsubsection{Minimal set of features}

With the data previously collected, it is possible to build a minimal set of features of different data sources, demographic, personality, academic and behavioral, mentioned earlier. It is not needed to have institutional data as it is a cross-population experiment. This minimal set was designed to be reproducible in other context and experiment and to make comparisons with other existing open data sets.

For the demographic category, several works \cite{Dutt2017, Adejo2018} show that basic demographic information on students have a significative role in characterizing students and predicting their success. Age, gender, level of education and field of education field are then kept. 

Moreover, time spent on tasks, number of clicks and whether an activity or resource has been viewed/done are commonly provided and used in similar experiments in the field~\cite{Feldman2015, HjAhmad2010, Popescu2009, SabineGraf2007}. Thus, such behavioral features are also part of the set.

Besides, past academic grades have a predictive importance, reflecting students’ consistency in their schooling \cite{Livieris2018, Adejo2018}. Course performance based on the grades obtained by participant before the final quiz constitute an academic history. Their average is independent of the number of obtained grades and is a reproducible feature that we used.

Regarding personality, few studies consider these kinds of students’ characteristics to explore correlation with their learning performance whereas \cite{Afzaal2019, Papadogiannis2020} recent works point out relevant relationships. Since composing a generic set becomes complex when including data for personality recognition, one motivation question and two participants’ text descriptions on themselves either positive or negative are the only information asked. The two latter will be processed in future work.

All this information constitutes the minimal set of features detailed in Table \ref{tab-minimal}. Demographic and personality data sources were collected through a questionnaire (see Appendix \ref{ap-questionnaire}) whereas all the other sources were automatically recorded.

\begin{table*}[!ht]
\caption{Minimal set of features. (Those which are marked with a * will be processed in future work).}
\label{tab-minimal}
\vskip 0.1in
\begin{minipage}{\textwidth} % <--- new for table footnotes
\begin{center}
\begin{small}
\begin{tabular}{llll}
\hline
Category                                                                    & Name           & Description                                                                                                                          & Values       \\ \hline
\multirow{4}{*}{\begin{tabular}[c]{@{}l@{}}Demographic \\ (D)\end{tabular}} & age            & Learner’s age                                                                                                                        & 18-99        \\
                                                                            & gender         & Learner’s gender                                                                                                                     & F/M/NA       \\
                                                                            & ed\_level      & Learner’s level of education (current or last)                                                                                       & 1-8\footnote{Number of years after high school graduation/A-level.}          \\
                                                                            & ed\_field      & Learner’s field of education (current or last)                                                                                          & 4 classes\footnote{Maths, Computer Science, Engineering, Technology... - Health, Biology, Medicine... - Humanities, Languages, Arts, Culture ... - Administration, Law, Management, Commerce...} \\ \hline
\begin{tabular}[c]{@{}l@{}}Academic \\ (A)\end{tabular}                     & avrg\_grade    & \begin{tabular}[c]{@{}l@{}}Average of the learner’s course grades before taking \\ the final exam\end{tabular}                       & 0.00-10.00   \\ \hline
\multirow{3}{*}{\begin{tabular}[c]{@{}l@{}}Behavioral \\ (B)\end{tabular}}  & \%\_completion & Percentage of course completion                                                                                                      & 0.00-1.00    \\
                                                                            & nb\_action     & \begin{tabular}[c]{@{}l@{}}Sum of the number of views on the resources and the \\ number of attempts at the activities\end{tabular} & 0-inf.       \\
                                                                            & time           & Total time learner spent on activities                                                                                               & 0-inf.       \\ \hline
\multirow{3}{*}{\begin{tabular}[c]{@{}l@{}}Personality \\ (P)\end{tabular}} & motivation     & Learner’s enrolment motivation                                                                                                       & 3 classes\footnote{Very interested in the course subject - Moderately interested but curious - Little interest but will to participate in the educational experience.}          \\
                                                                            & descr\_pos*    & Learner’s description of a positive day                                                                                              & Text         \\
                                                                            & descr\_neg*    & Learner’s description of a negative day                                                                                              & Text         \\ \hline
\end{tabular}
\end{small}
\end{center}
\end{minipage} % <--- new
\vskip -0.2in
\end{table*}

\subsubsection{Additional features}

Additional features were created to explore other hypothetical relationships (see Table \ref{tab-additional}). These new features average the grades grouped by format type or content type of the learning material and aim at discovering relationships between hypothetical learning preferences and students’ performance \cite{Antoniuk2019}. These are visual or verbal preference regarding the format type and factual or practical regarding the content type. Two other features correspond to a memory and a deduction score. Related analyses are presented in the results section.

\begin{table*}
\caption{Additional features.}
\label{tab-additional}
\vskip 0.1in
\begin{center}
\begin{small}
\begin{tabular}{llll}
\hline
Category                                                                                & Name      & Description                                           & Values    \\ \hline
\multirow{6}{*}{\begin{tabular}[c]{@{}l@{}}Learning \\ preferences \\ (L)\end{tabular}} & visual    & Learner’s score on visual material                    & 0.00-1.00 \\
                                                                                        & verbal    & Learner’s score on written material                   & 0.00-1.00 \\
                                                                                        & factual   & Learner’s score on factual content            & 0.00-1.00 \\
                                                                                        & practical & Learner’s score on practical content          & 0.00-1.00 \\
                                                                                        & memory    & Learner’s score on memorizing any kind of information & 0.00-1.00 \\
                                                                                        & deduction & Learner’s score on deducing any kind of knowledge     & 0.00-1.00 \\ \hline
\end{tabular}
\end{small}
\end{center}
\vskip -0.2in
\end{table*}

To have visual insights on all the features, both minimal and additional, please see Appendix \ref{ap-viz}.

\subsection{Supervised learning algorithms}

For the sake of explainability and understanding on key factors affecting students’ performance to answer the research questions, the requirements of a white-box supervised learning algorithm were met. Features should be understandable and the machine learning process transparent \cite{Loyola-Gonzalez2019}.

On this point, decision tree and rule-based learning algorithms are supported for the students’ performance prediction task in the systematic review of \cite{Alamri2021}. Decision trees are indeed self-explanatory and deal with any type of features \cite{Breiman1984}. On the other hand, they are sensitive to small variations in the training dataset \cite{Quinlan1986}.

In this work, the prediction results were obtained using a CART decision tree algorithm\footnote{\url{https://scikit-learn.org/stable/modules/tree.html\#tree}}. In some of the following experiments, they were also compared with a random forest (RF) algorithm and a support vector machine (SVM) algorithm to evaluate the relative performance of the decision tree (DT). These algorithms’ parameters were selected by performing a grid search for each experiment and by considering the computational resources available.

\subsection{Feature selection algorithms}

To assess the relevance of the features, we used feature selection algorithms, three of which were forward elimination (FE), backward elimination (BE) and recursive feature elimination (RFE). FE, also called forward selection\footnote{\url{https://scikit-learn.org/stable/modules/feature_selection.html\#sequential-feature-selection}}, is a greedy procedure that iteratively finds the best new feature to add to the set of selected features. We start with zero feature and at each iteration we keep adding the feature which best improves our model. The procedure stops when an addition of a new feature does not improve the performance of the model or when the desired number of selected features is reached. BE, also called backward selection, works in the opposite direction to FE. We thus start with all the features and greedily remove the least significant feature at each iteration. Then, RFE\footnote{\url{https://scikit-learn.org/stable/modules/generated/sklearn.feature_selection.RFE.html}} is an instance of BE \cite{Guyon2002} that creates models and determines the worst performing feature at each iteration to find the best performing feature subset based on the feature ranking. Finally, these wrapper methods were used with our decision tree models.

\section{Results}

\begin{table*}
\begin{center}
\caption{Comparison of datasets. (D: demographic. A: academic. B: behavioral. P: personality. L: learning preferences).}
\vskip 0.1in
\label{tab-datasets}
\begin{small}
\begin{tabular}{l|ccc}
\hline
\multicolumn{1}{c|}{\textbf{}} & D1                                                           & D2                                                           & D3                                                              \\ \hline
\# courses                     & \begin{tabular}[c]{@{}c@{}}1 \\ (general topic)\end{tabular} & \begin{tabular}[c]{@{}c@{}}7 \\ (3 disciplines)\end{tabular} & \begin{tabular}[c]{@{}c@{}}238 \\ (10 disciplines)\end{tabular} \\
\# students                    & 44                                                           & 28,785                                                       & 224,914                                                         \\
\# training instances          & 27                                                           & 11,420                                                       & 10,005                                                          \\
class-imbalance ratio          & 59.25\%                                                      & 72.52\%                                                      & 36.78\%                                                         \\
data sources                   & D + B + A + P + L                                            & D + B + A                                                    & D + B                                                           \\ \hline
\end{tabular}
\end{small}
\end{center}
\vskip -0.2in
\end{table*}

Three datasets were considered and named respectively D1, D2 and D3. D1 is built from the collected data described in section \ref{methods} where 44 students were involved. D2 is the Open University Learning Analytics dataset \cite{J.2017} which can be built from several data tables\footnote{\url{https://analyse.kmi.open.ac.uk/open_dataset\#description}}. It has 28,785 students involved and 7 possible courses which could have been taken at defined starting dates in the years 2013 and 2014. D3 is the Canvas Network Person dataset \cite{Network2016} comprised of de-identified data from Canvas Network open courses running between January 2014 and September 2015. It contains records from 224,914 students enrolled in 238 possible courses belonging to 10 different disciplines. A summary is displayed in Table \ref{tab-datasets}.

As well as D1, the two datasets D2 and D3 are about varied students’ populations. Nonetheless, the two latter were not designed for multiple data sources. After some cleaning and preprocessing to make fair comparisons between the three datasets, 27, 11,420 and 10,005 instances were kept from D1, D2 and D3 respectively. It can be noted that the gender attribute was used for data understanding but removed for prediction to be gender inclusive with respect to the predictions made. We recall that the target variable is binary (pass (1) or fail (0)).

\subsection{Comparison with open datasets}

The first experiment aims at positioning D1 with common educational data represented by D2 and D3. As there is no standard approach yet to collect educational data, students’ age, level of education and number of interactions were the only common features found across the three datasets. D3 is indeed the most restrictive. These features represent demographic (D) and behavioral (B) data.

We computed baseline results evaluated with ten-fold cross-validation and we compared them with four tests performing different balancing techniques: upsampling the minority class, downsampling the majority class, up and downsampling both classes and Synthetic Minority Oversampling Technique (SMOTE \cite{Chawla2002}). The best results are presented in Table \ref{fig-best_balancing} (and the others in Appendix \ref{ap-balancing}). Across the different balancing technique tests, these techniques especially improved performance with D1 whereas D2 and D3 showed more stable results and always had close performance. At the end, upsampling the minority class and downsampling the majority class presented the best results globally and will be reused for all the other experiments. 

\begin{table}[!ht]
\begin{center}
\caption{Results with the upsampled minority class and the downsampled majority class.}
\vskip 0.1in
\label{fig-best_balancing}
\begin{small}
\begin{tabular}{l|lll}
\hline
               & D1    & D2    & D3    \\ \hline
Accuracy (\%)  & 66.66 & 69.05 & 69.77 \\
Precision (\%) & 59.16 & 69.15 & 70.06 \\
Recall (\%)    & 67.50 & 69.05 & 69.77 \\
F-score (\%)   & 57.99 & 69.01 & 69.66 \\ \hline
\end{tabular}
\end{small}
\end{center}
\vskip -0.2in
\end{table}

\begin{figure}[ht]
\vskip 0.1in
\begin{center}
\centerline{\includegraphics[scale=0.6]{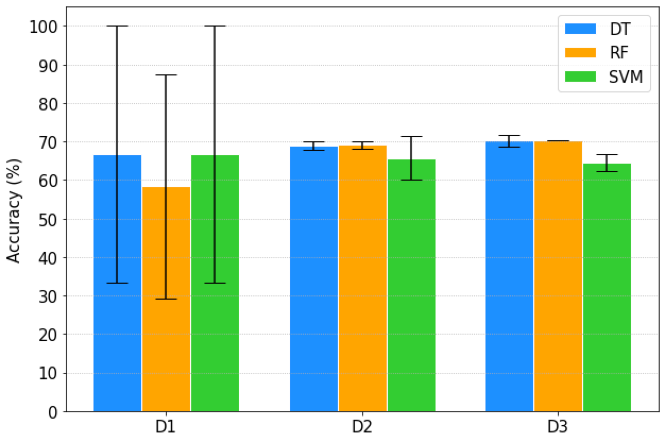}}
\caption{Resulting accuracies with DT, RF and SVM on D1, D2 and D3.}
\label{fig-exp1}
\end{center}
\vskip -0.2in
\end{figure}

We then display in Figure \ref{fig-exp1} the predictive performances with error bars obtained on the three datasets with the different models. Not surprisingly, D1 leads to lower performance and higher variability than D2 and D3 since D1 has comparatively much less instances. It corresponds to what we expected but the mean accuracy is encouraging since, with more samples, it is probable to obtain closer results. Furthermore, the DT model seems relevant for predictive performance compared to the other models, in addition to its explainability.

Besides, we also experimented transfer learning between the datasets to see if there is a presumably shared knowledge with open educational data and our collected data. The results are in Figures \ref{fig_tl_dt}, \ref{fig_tl_rf} and \ref{fig_tl_svm} in Appendix \ref{ap-transfer}. We deduced encouraging results again as in the three graphs, when testing on D1, we obtained about the same performance when we trained our DT model either on D2 or D3 than when we trained the model on D1 (normal case).

\subsection{Usual data sources }

\begin{figure}[ht]
\vskip 0.1in
\begin{center}
\centerline{\includegraphics[scale=0.5]{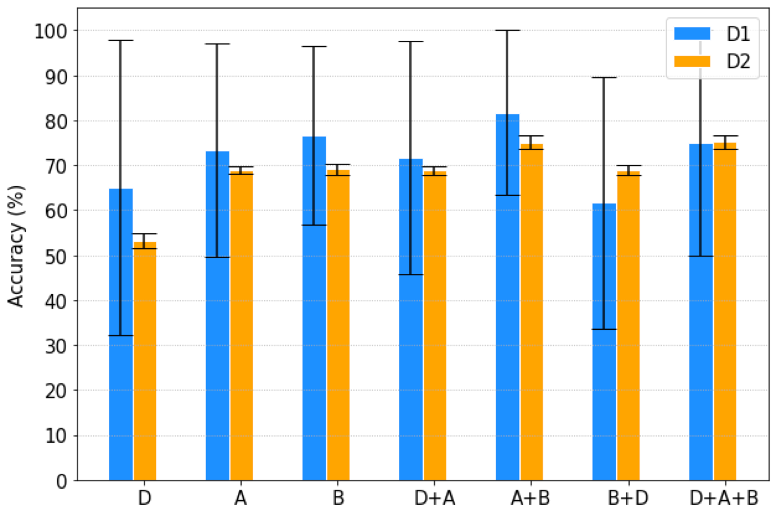}}
\caption{Resulting accuracies with the different usual sources.}
\label{fig-exp2}
\end{center}
\vskip -0.2in
\end{figure}

The second experiment aims at comparing observable trends between the collected data D1 and the open data D2, that represent different populations and courses, regarding the usual demographic (D), academic (A) and behavioral (B) data sources (see Appendix \ref{ap-exp2_features} for details on the features). D3 was excluded since it has too few comparable features. The experiment consists of comparing performance results on different feature combinations. The results are in Figure \ref{fig-exp2} and comments in terms of performance and feature importance are below. The detailed results are in Table \ref{tab-exp2} in Appendix \ref{ap-exp2_rez}.

Academic data and behavioral data enable to perform the best results with both datasets. They are indeed the features most representative of what is being predicted, students’ performance, which supports their common use. When combined, results slightly improve, with a stronger emphasis with D2. We also observe the same ratio about 40-60\% of prediction importance between academic data and behavioral data within the two datasets. Besides, the time is the main predictor among the behavioral data.

Regarding demographic data, it shows one of the lowest performances when compared to other combinations. In fact, these demographic features are too general to discriminate groups of students (see distributions according the target in Appendix \ref{ap-scatter}) and other features such as working conditions or socio-economic features (work and living environments, parental education, financial background, etc.) could be more useful. In addition, about the same performances are obtained when demographic data are combined with other sources like academic or behavioural data, suggesting that it does not bring additional information and even add noise in some cases with lower results. Feature importance confirms that academic features are 99\% and 100\% important in D2 and D1 respectively, and that behavioral features in particular the time are 98\% important too, compared to demographic ones (0-2\%). However, these general demographic features are very useful to understand the data and make statistical analyses outside a prediction task so that they are worth collecting every time an educational study is performed.

To conclude this part, we observe similar trends in both datasets representing different students’ populations and courses. Demographic data has low prediction importance contrary to academic data and behavioral data. Moreover, a combination of the two latter gives the best results. Thus, using only two of the usual data sources seems to be relevant.

\subsection{Additional data sources}

The third experiment aims at testing and comparing additional combinations provided by D1. The same approach as the previous subsection is adopted and the results tables and figures are displayed in Appendix \ref{ap-additional}. 

For the usual data sources, we obviously find the same conclusions than the previous experiment including the 40-60\% ratio between academic and behavioral data when combined, and the time as a dominant behavioral feature (87\% against 13\% for the number of interactions since the percentage of completion has a null variance in D1). Besides, demographic data still has a quasi-null or null importance when combined with any type of data, and academic and behavioral data always have high importance (90-100\%) when not in the same combination (for instance with personality data). Nonetheless, learning preferences data that are derived from academic data show only 36\% of importance for the verbal score which is the most significant learning preferences feature. It suggests that learning preferences could be useful for prediction but not in every case like academic data. Among the learning preferences features, the verbal, the factual, visual and deduction scores have a consistent performance across the several-source combinations with a more significative role of the verbal score. Then, a general performance loss is observed with 3 sources or more especially with 5 sources which give the lowest performances.

To conclude this part, the level of importance either of demographic data, academic data or behavioral data generalizes with additional data sources, except in the case when academic data are combined with learning preferences data as the latter is derived from the former. The time is still the most important predictor among behavioral data and some scores, in particular the verbal one, seem to be significantly important. Finally, testing 1-source, 2-source, 3-source, 4-source and 5-source combinations leads to consider 2-source combinations as the most performing combinations such as academic and behavioral data again.

\subsection{Feature selection analyses}
\label{fs}

\begin{figure*}[t]
\vskip 0.1in
  \centering
  \subfigure[FE]{\includegraphics[scale=0.6]{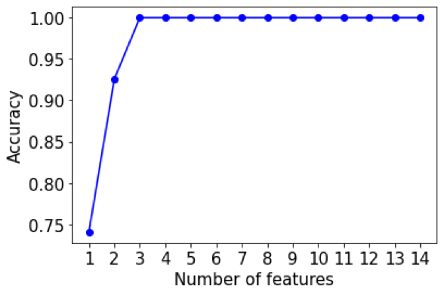}}\quad
  \hfill
  \subfigure[BE]{\includegraphics[scale=0.6]{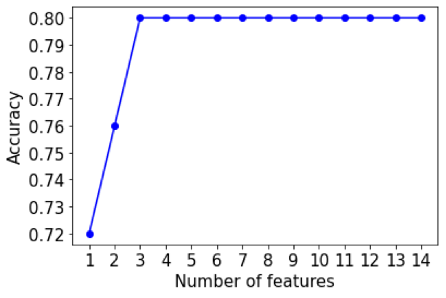}}
  \hfill
  \subfigure[RFE with cross-validation]{\includegraphics[scale=0.32]{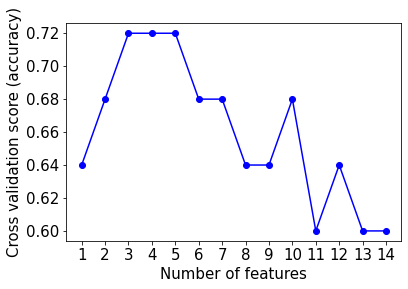}}
  \caption{Feature selection wrapper methods results.}
  \label{fig-wrappers}
  \vskip -0.2in
\end{figure*}

In this part, we performed feature selection in order to assess the relevance of the features individually in contrast to the previous combinations. All the features in the minimal set and the additional set are considered. As there are many features and comparably few instances, feature selection leads to simplifying the model for better understanding and improving prediction performance. Reducing computational cost and time is not the main purpose but this outcome will be also met. We first used wrapper methods and then filter methods to first find an optimal feature set for the prediction task and then to have insights on the relationships between the features for domain knowledge (usefulness vs. relevance \cite{Guyon2003}). It has to be noted that the feature of percentage of course completion has a null variance and was not taken into account for the following since the preprocessing step kept only records of students who fully completed the course, including filling the questionnaire, to have all the needed data notably the related demographic one for the study.

Forward elimination (FE), backward elimination (BE) and recursive feature elimination (RFE) with cross-validation are performed to find a smaller set of not redundant or complementary features. According to the experimental results, the optimal number of features appears to be 3, corresponding mostly to a 2- or 3-source combination with one or few features in each source. Taking 3 as the optimal number of features, the selected ones from each of the methods are indeed: time (B), verbal score (L) and deduction score (L) for RFE; field of education (D), time (B) and verbal score (L) for FE; and the same as FE for BE.
%\begin{itemize}[noitemsep, nolistsep]
%    \item RFE: time (B), verbal score (L) and deduction score (L),
%    \item FE: learner’s field of education (D), time (B), verbal score (L),
%    \item BE: same as FE.
%\end{itemize}
These results emphasize the predictive power of the time and the verbal score whose importance was demonstrated previously. 

Regarding filter methods, variable ranking methods were used such as statistical measures ANOVA (linear correlation coefficients) and Kendall (nonlinear correlation coefficients). The correlation between each feature and the target are computed in Figure \ref{fig-anova} and Figure \ref{fig-kendall}. Thus, independently of the prediction model, the verbal score is again significantly relevant for students’ performance. Academic data or learning preferences data derived from the former are also well ranked, followed by behavioral data, demographic data and personality data. Among the behavioral features, the time is indeed better ranked than the number of actions and among the demographic data, learner’s field of education has a much higher correlation with the target. This can be easily explained by the prior knowledge brought by the field of education that helps the learner to better succeed. Besides, in the population of D1, learner’s level of education and learner’s age are highly correlated as seen in Figure \ref{fig-corr_matrix}, describing why a good coefficient of one of them implies a low coefficient for the other in Figure \ref{fig-anova} and Figure \ref{fig-kendall}.

To conclude this part, this feature selection study, combined with the previous experiments, stresses the usefulness and the relevance of the verbal score and the time. Such role for the verbal score means that the skills of well extract, memorize and deduce knowledge from written learning material could be a strong advantage and a major ability related to success. 
%It indeed matches with how well a student, compared to others, is able to learn from most of the knowledge taught with written explanations, textbooks, etc. 
However, the statistical measures expose more general relationships such as students’ field of education or average grade, which allows to have a more holistic view of students contrary to the time for instance. Moreover, the wrapper methods point out again that 2 data sources or about 3 features lead to the best results with a minimal and complementary set of features. Academic data and behavioral data are still the preferred sources.

\section{Conclusion}

In this work, we predicted students’ performance using multiple data sources from online courses, including one we created to that purpose. We compared results with different learning algorithms and different datasets. Two research questions were posed: (1) To what extent using multiple data sources on learners increases prediction performance? (2) Which learners’ characteristics are most related to higher academic performance? The results showed encouraging similar trends and knowledge with different students’ populations and courses. We highlighted the relevance and the importance of academic and behavioral data in contrast to demographic ones and the use of multiple data sources where the optimal number of complementary features, hence complementary data on students', appears to be 3. Eventually, we formulated a hypothetical advantageous relationship between students’ who have stronger abilities of memorizing and deducing knowledge from written material and better performance. However, it has to be kept in mind that most of the results are based on a small number of instances stressing the limitations of the study.

Future work will be related to having more participants to the learning experiment to challenge these current results, implementing a personality traits recognition process to enrich personality features, exploiting the explainability of the decision trees, creating new features from existing ones, and approaching the prediction task as a multi-class classification problem for refined results.

% Note use of \abovespace and \belowspace to get reasonable spacing
% above and below tabular lines.

%\begin{table}[t]
%\caption{Classification accuracies for naive Bayes and flexible
%Bayes on various data sets.}
%\label{sample-table}
%\vskip 0.1in
%\begin{center}
%\begin{small}
%\begin{sc}
%\begin{tabular}{lcccr}
%\toprule
%Data set & Naive & Flexible & Better? \\
%\midrule
%Breast    & 95.9$\pm$ 0.2& 96.7$\pm$ 0.2& $\surd$ \\
%Cleveland & 83.3$\pm$ 0.6& 80.0$\pm$ 0.6& $\times$\\
%Glass2    & 61.9$\pm$ 1.4& 83.8$\pm$ 0.7& $\surd$ \\
%Credit    & 74.8$\pm$ 0.5& 78.3$\pm$ 0.6&         \\
%Horse     & 73.3$\pm$ 0.9& 69.7$\pm$ 1.0& $\times$\\
%Meta      & 67.1$\pm$ 0.6& 76.5$\pm$ 0.5& $\surd$ \\
%Pima      & 75.1$\pm$ 0.6& 73.9$\pm$ 0.5&         \\
%Vehicle   & 44.9$\pm$ 0.6& 61.5$\pm$ 0.4& $\surd$ \\
%\bottomrule
%\end{tabular}
%\end{sc}
%\end{small}
%\end{center}
%\vskip -0.1in
%\end{table}

\section*{Acknowledgements}

We would like to express our deepest appreciation to all those who contributed to this project. We would like to acknowledge the support of The Open University who gave their permission to use the required features on their e-learning platform. We would also like to thank our main advisors and collaborators, especially Isabelle Guyon, Marc Evrard, the course reviewers, including Alice Lacan who also helped with the communication of the learning experiment, and finally, all the volunteer participants.


\bibliography{main.bib}
\bibliographystyle{icml2021}


\newpage

\appendix

\renewcommand{\textfraction}{.01}

\section{Questionnaire for collecting demographic and personality data}
\label{ap-questionnaire}

The main course and questionnaire were in French for communication and spread purpose of the learning experiment. We thus report this questionnaire version with translations.

1. Vous êtes (\textit{You are}) :
\begin{itemize}[noitemsep, nolistsep]
    \item une femme (\textit{a woman}).
    \item un homme (\textit{a man}).
    \item Ne se prononce pas (\textit{no answer}).
\end{itemize}

2. Quel âge avez-vous ? \textit{How old are you?}
\begin{itemize}[noitemsep, nolistsep]
    \item Réponse numérique. \textit{Numerical answer}.
\end{itemize}

3. Le français est-il votre langue maternelle ? \textit{Is French your native language?}
\begin{itemize}[noitemsep, nolistsep]
    \item Oui (\textit{yes}).
    \item Non (\textit{no}).
\end{itemize}

4. Quel est votre niveau d’études actuel (ou dernier) ? \textit{What is your current (or last) level of education?}
\begin{itemize}[noitemsep, nolistsep]
    \item Choix de Bac +1 à Bac +8. \textit{Categorical answer from A-level to PhD}.
\end{itemize}

5. Dans quelle filière d’études êtes-vous (ou la dernière) ? \textit{What is your current (or last) field of education?}
\begin{itemize}[noitemsep, nolistsep]
    \item Maths, Informatique, Ingénierie, Technologie... \textit{Maths, Computer Science, Engineering, Technology}...
    \item Santé, Biologie, Médecine... \textit{Health, Biology, Medicine}...
    \item Lettres, Langues, Arts, Culture... \textit{Humanities, Languages, Arts, Culture}...
    \item Administration, Droit, Management, Commerce... \textit{Administration, Law, Management, Commerce}...
\end{itemize}

6. En apprendre plus sur le cerveau vous intéresse :
\begin{itemize}[noitemsep, nolistsep]
    \item beaucoup (le sujet m’intéresse). \textit{You are very interested in learning more about the brain (you are interested in the subject)}.
    \item moyennement (mais je suis curieux(se)). \textit{You are moderately interested in learning more about the brain (but you are curious)}.
    \item peu (mais je souhaite participer à cette expérience pédagogique). \textit{You have little interest in learning more about the brain (but you want to participate in this educational experience)}.
\end{itemize}

7. Racontez une histoire hypothétique dans laquelle vous décrivez une journée idéale où tout s'est bien passé. Exemple : Aujourd'hui était une super journée pour moi, je me suis réveillé(e) tôt le matin et j'ai pu courir 5 km avant de me rendre au travail. Pendant que je travaillais, mon ami George m'a apporté une tasse de café, c'était une bonne surprise car la machine à café du bureau ne fonctionne pas. J'ai déjeuné avec Lucy, George et Mike, c'était vraiment sympa. Le soir, j'ai pu sortir mon chien pour une petite promenade et j'ai eu une belle vue sur la lune. \textit{Tell a hypothetical story in which you describe an ideal day when everything went well. Example: Today was a great day for me, I woke up early in the morning and was able to run 5 km before going to work. While I was working, my friend George brought me a cup of coffee, which was a nice surprise because the office coffee machine doesn't work. I had lunch with Lucy, George and Mike, which was really nice. In the evening I was able to take my dog out for a little walk and had a nice view of the moon.}
\begin{itemize}[noitemsep, nolistsep]
    \item Description libre. \textit{Free description}.
\end{itemize}

8. Racontez une histoire hypothétique dans laquelle vous décrivez une mauvaise journée où vous avez rencontré des difficultés. Exemple : Aujourd'hui n'était pas une bonne journée pour moi, je me sentais très fatigué(e) le matin et il m'était difficile de me concentrer au travail le matin. Les choses se sont améliorées pendant le déjeuner, j'ai pu discuter un peu avec Lucy et elle m'a invité à une fête ce week-end. Cependant, il pleuvait quand j'ai quitté le bureau et j'ai oublié mon parapluie, j'étais vraiment trempé(e) quand je suis arrivé(e) à la maison, à cause du temps je n'ai pas pu sortir mon chien aujourd'hui, j'espère que je pourrai le faire demain. \textit{Tell a hypothetical story in which you describe a bad day when you had difficulties. Example: Today was not a good day for me, I felt very tired in the morning and it was difficult to concentrate at work in the morning. Things improved over lunch, I was able to chat to Lucy a bit and she invited me to a party at the weekend. However, it was raining when I left the office and I forgot my umbrella, I was really soaked when I got home, because of the weather I couldn't take my dog out today, I hope I can do it tomorrow.}
\begin{itemize}[noitemsep, nolistsep]
    \item Description libre. \textit{Free description}.
\end{itemize}

9. J'accepte en suivant ce cours de donner la permission d'utiliser mes réponses anonymes pour une étude scientifique sur les modèles d'apprentissage en ligne. Aucune donnée personnelle ne sera stockée ni partagée (plus d'informations ici\footnote{\url{https://sites.google.com/view/braincourse/utilisation-des-donnees}}). \textit{By taking this course I consent to give permission for the usage of my anonymized answers for a scientific study associated with e-learning patterns. Your personal data will not be stored or shared (more information here).}
\begin{itemize}[noitemsep, nolistsep]
    \item Oui (\textit{yes}).
\end{itemize}

\newpage
\section{Data visualization (collected data)}
\label{ap-viz}

\begin{figure}[ht]
\vskip 0.1in
\begin{center}
\centerline{\includegraphics[scale=0.4]{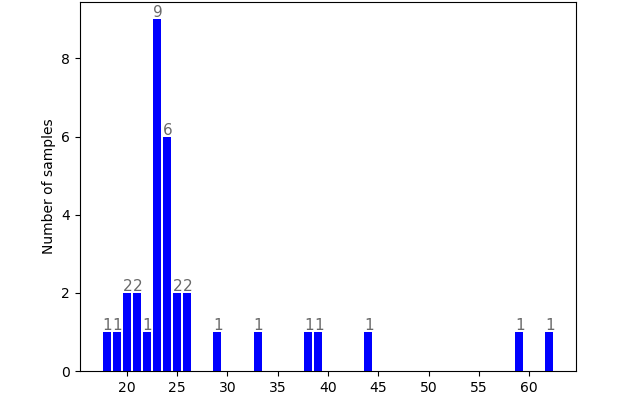}}
\caption{Learners' age.}
\end{center}
\vskip -0.1in
\end{figure}

\begin{figure}[ht]
\vskip 0.1in
\begin{center}
\centerline{\includegraphics[scale=0.4]{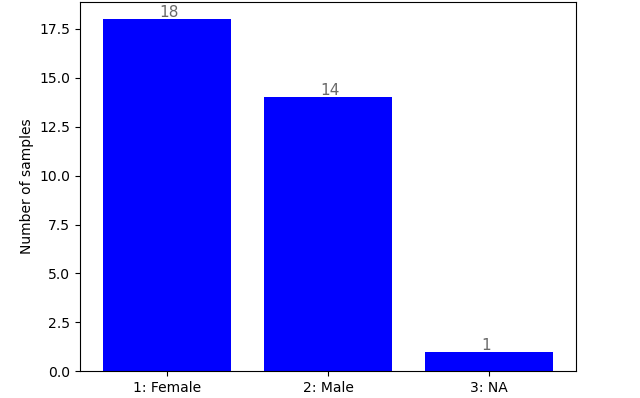}}
\caption{Learners' gender.}
\end{center}
\vskip -0.1in
\end{figure}

\begin{figure}[ht]
\vskip 0.1in
\begin{center}
\centerline{\includegraphics[scale=0.4]{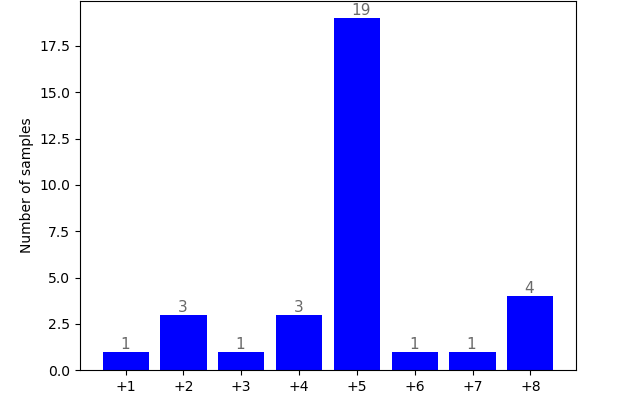}}
\caption{Learners' level of education.}
\end{center}
\vskip -0.1in
\end{figure}

\begin{figure}[!ht]
\vskip 0.1in
\begin{center}
\centerline{\includegraphics[scale=0.4]{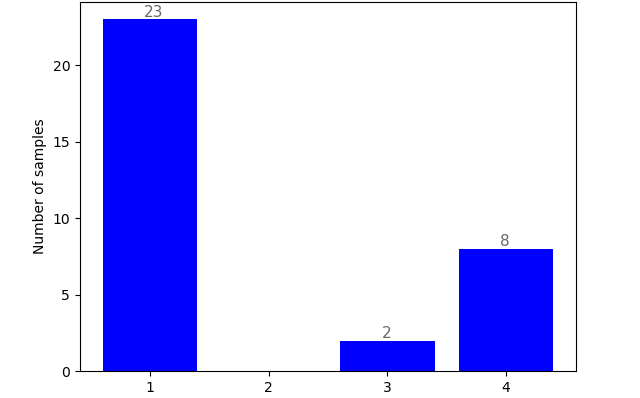}}
\caption{Learners' field of education.}
\end{center}
\vskip -0.1in
\end{figure}

\begin{figure}[!ht]
\vskip 0.1in
\begin{center}
\centerline{\includegraphics[scale=0.4]{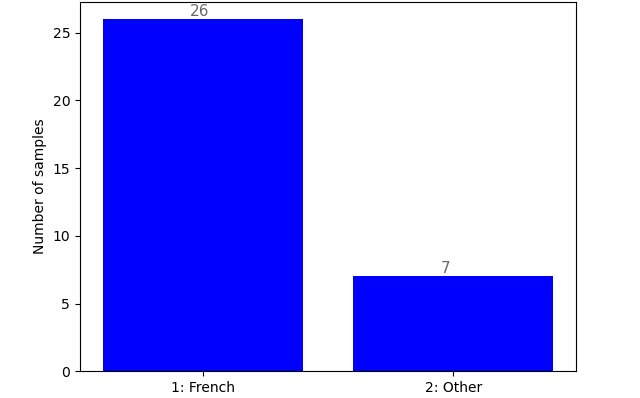}}
\caption{Learners' native language.}
\end{center}
\vskip -0.1in
\end{figure}

\begin{figure}[!ht]
\vskip 0.1in
\begin{center}
\centerline{\includegraphics[scale=0.4]{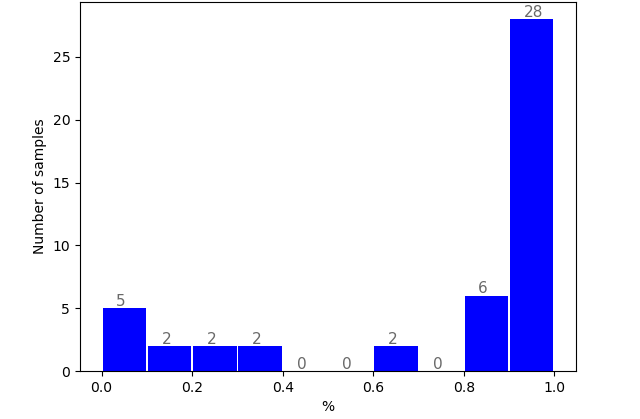}}
\caption{Learners' percentage of completion.}
\end{center}
\vskip -0.1in
\end{figure}

\begin{figure}[ht]
\vskip 0.1in
\begin{center}
\centerline{\includegraphics[scale=0.4]{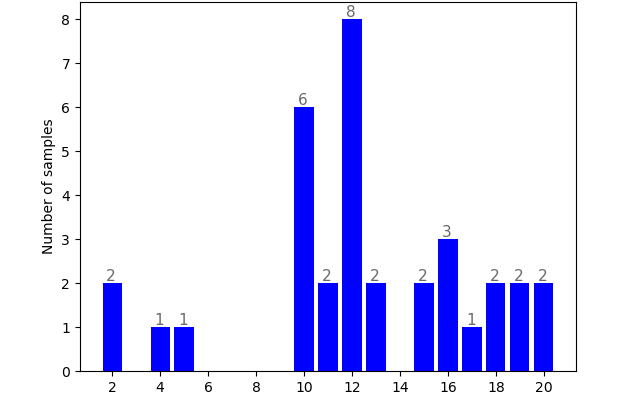}}
\caption{Learners' number of actions.}
\end{center}
\vskip -0.1in
\end{figure}

\clearpage
\begin{figure}[ht]
\vskip 0.1in
\begin{center}
\centerline{\includegraphics[scale=0.4]{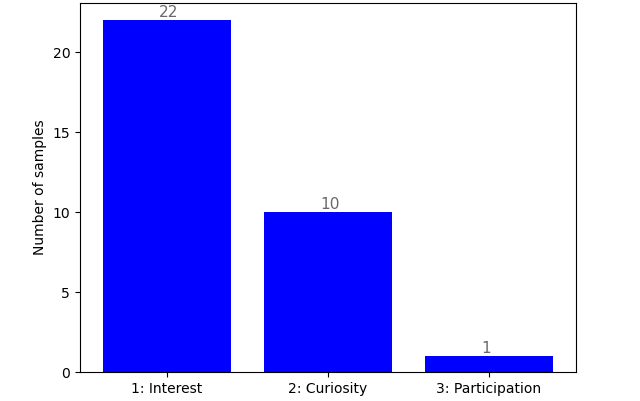}}
\caption{Learners' motivation.}
\end{center}
\vskip -0.1in
\end{figure}

\begin{figure}[ht]
\vskip 0.1in
\begin{center}
\centerline{\includegraphics[scale=0.4]{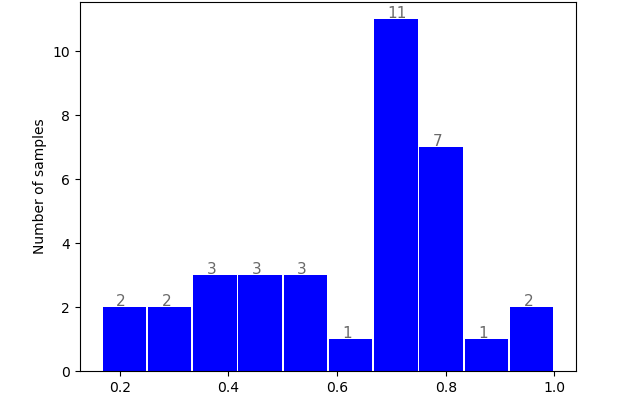}}
\caption{Learners' visual score.}
\end{center}
\vskip -0.1in
\end{figure}

\begin{figure}[!h]
\vskip 0.1in
\begin{center}
\centerline{\includegraphics[scale=0.4]{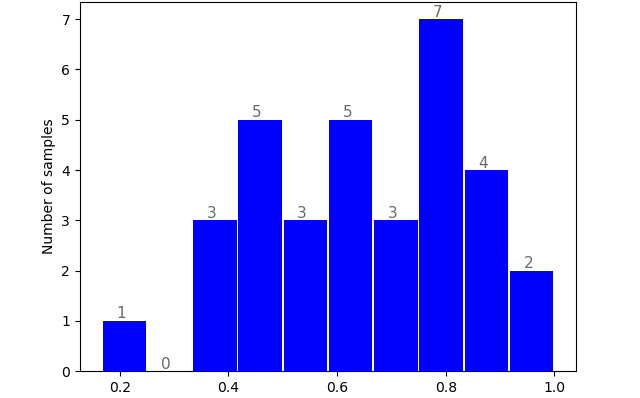}}
\caption{Learners' verbal score.}
\end{center}
\vskip -0.1in
\end{figure}

\begin{figure}[ht]
\vskip 0.1in
\begin{center}
\centerline{\includegraphics[scale=0.4]{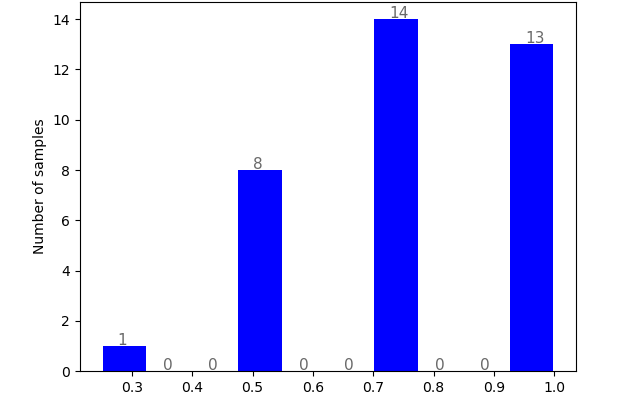}}
\caption{Learners' factual score.}
\end{center}
\vskip -0.1in
\end{figure}

\begin{figure}[!ht]
\vskip 0.1in
\begin{center}
\centerline{\includegraphics[scale=0.4]{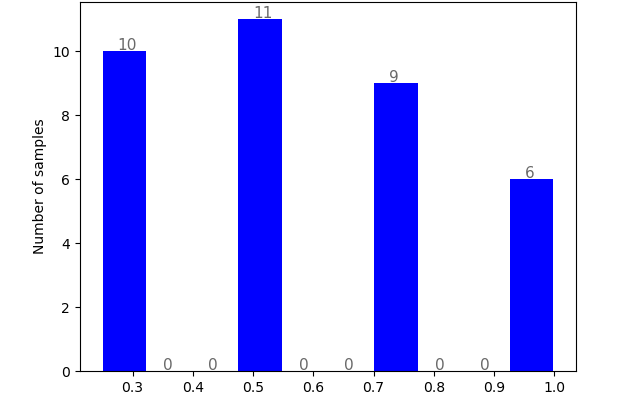}}
\caption{Learners' practical score.}
\end{center}
\vskip -0.1in
\end{figure}

\begin{figure}[ht]
\vskip 0.1in
\begin{center}
\centerline{\includegraphics[scale=0.4]{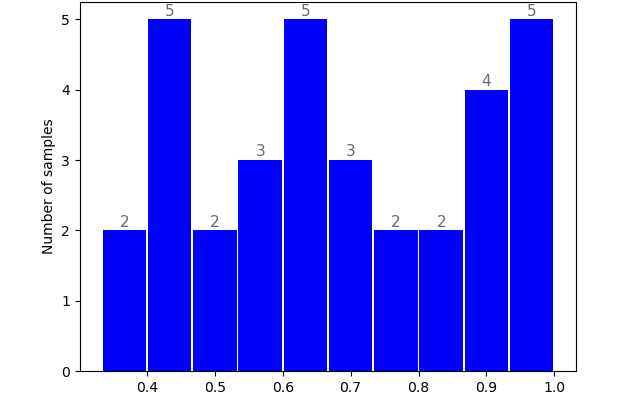}}
\caption{Learners' memory score.}
\end{center}
\vskip -0.1in
\end{figure}

\begin{figure}[ht]
\vskip 0.1in
\begin{center}
\centerline{\includegraphics[scale=0.4]{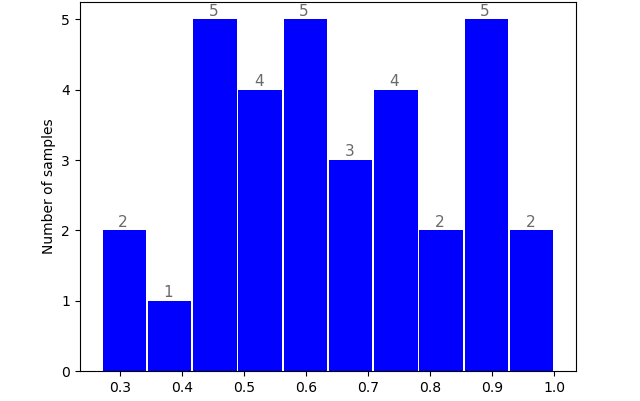}}
\caption{Learners' deduction score.}
\end{center}
\vskip -0.1in
\end{figure}

\clearpage
\newpage
\begin{figure}
\vskip 0.1in
     \centering
     \hfill
     \begin{subfigure}
         \centering
         \includegraphics[width=\columnwidth]{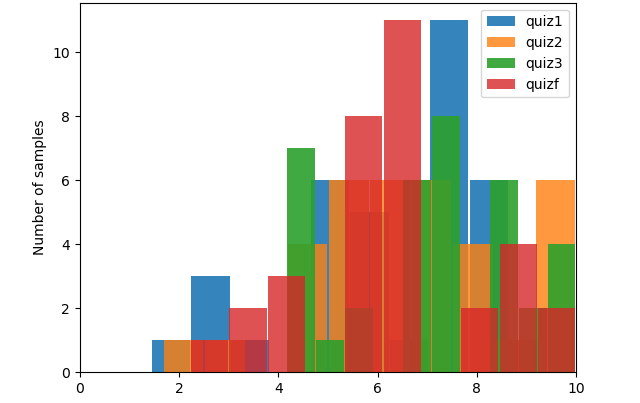}
     \end{subfigure}
     \hfill
     \begin{subfigure}
         \centering
         \includegraphics[width=\columnwidth]{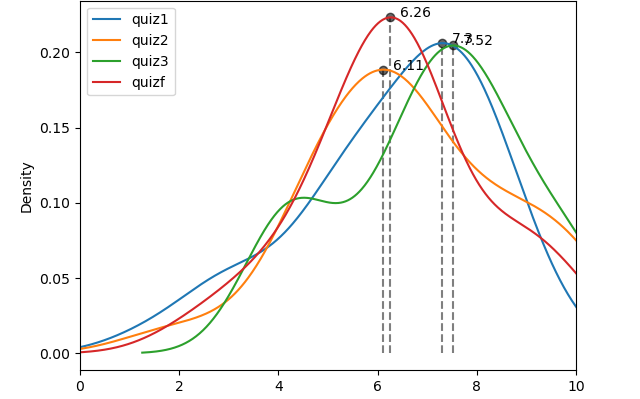}
     \end{subfigure}
        \caption{Learners' quiz grades and their kernel density estimates.}
\vskip -0.1in
\end{figure}

\begin{figure}[t]
\vskip 0.1in
     \centering
     \begin{subfigure}
         \centering
         \includegraphics[width=\columnwidth]{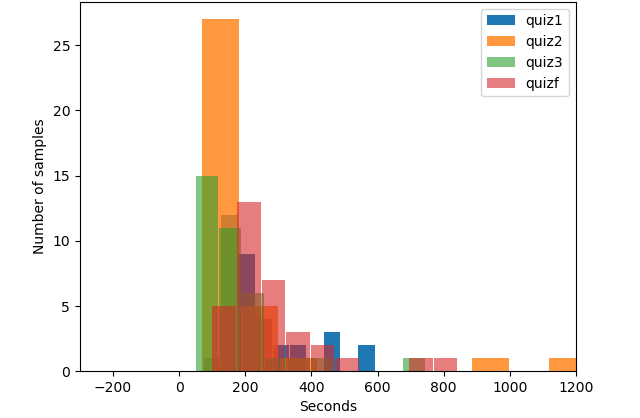}
     \end{subfigure}
     \begin{subfigure}
         \centering
         \includegraphics[width=\columnwidth]{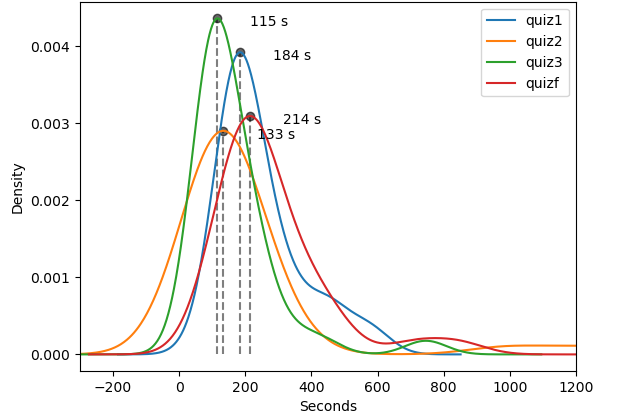}
     \end{subfigure}
        \caption{Learners' quiz times and their kernel density estimates.}
\vskip -0.1in
\end{figure}

\newpage
~
\newpage
~
\newpage
\section{Balancing techniques results}
\label{ap-balancing}

\begin{table}[!h]
\begin{center}
\caption{Baseline results without balancing technique.}
\vskip 0.1in
\begin{tabular}{l|lll}
\hline
                                    & D1    & D2    & D3    \\ \hline
\multicolumn{1}{l|}{Accuracy (\%)}  & 48.33 & 77.63 & 69.34 \\
\multicolumn{1}{l|}{Precision (\%)} & 42.50 & 78.25 & 67.28 \\
\multicolumn{1}{l|}{Recall (\%)}    & 50.00 & 65.09 & 67.41 \\
\multicolumn{1}{l|}{F-score (\%)}   & 41.66 & 65.85 & 67.28 \\ \hline
\end{tabular}
\vskip -0.1in
\end{center}
\end{table}

\begin{table}[!h]
\begin{center}
\caption{Results with upsamling.}
\vskip 0.15in
\begin{tabular}{l|lll}
\hline
               & D1    & D2    & D3    \\ \hline
Accuracy (\%)  & 52.49 & 69.81 & 69.72 \\
Precision (\%) & 47.50 & 70.09 & 69.95 \\
Recall (\%)    & 55.00 & 69.82 & 69.72 \\
F-score (\%)   & 47.16 & 69.71 & 69.63 \\ \hline
\end{tabular}
\vskip -0.1in
\end{center}
\end{table}

\begin{table}[!h]
\begin{center}
\caption{Results with downsamling.}
\vskip 0.1in
\begin{tabular}{l|lll}
\hline
          & D1    & D2    & D3    \\ \hline
Accuracy (\%)  & 66.66 & 69.82 & 69.49 \\
Precision (\%) & 53.33 & 70.18 & 69.70 \\
Recall (\%)    & 65.00 & 69.82 & 69.49 \\
F-score (\%)   & 57.33 & 69.69 & 69.41 \\ \hline
\end{tabular}
\vskip -0.1in
\end{center}
\end{table}

\begin{table}[!h]
\begin{center}
\caption{Results with SMOTE.}
\vskip 0.1in
\begin{tabular}{l|lll}
\hline
               & D1    & D2    & D3    \\ \hline
Accuracy (\%)  & 59.16 & 71.19 & 67.60 \\
Precision (\%) & 59.16 & 71.92 & 67.97 \\
Recall (\%)    & 60.0  & 71.19 & 67.60 \\
F-score (\%)   & 55.49 & 71.02 & 67.45 \\ \hline
\end{tabular}
\vskip -0.1in
\end{center}
\end{table}

\newpage
\section{Transfer learning results}
\label{ap-transfer}

\begin{figure}[ht]
\vskip 0.1in
\begin{center}
\centerline{\includegraphics[width=\columnwidth]{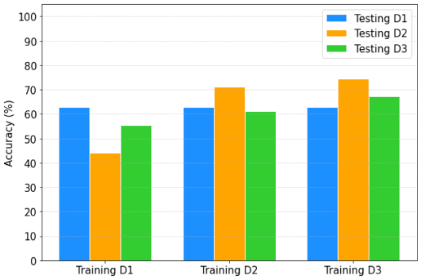}}
\caption{Results with DT.}
\label{fig_tl_dt}
\end{center}
\vskip -0.1in
\end{figure}

\begin{figure}[ht]
\vskip 0.1in
\begin{center}
\centerline{\includegraphics[width=\columnwidth]{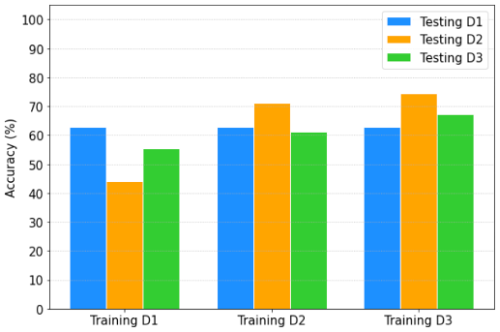}}
\caption{Results with RF.}
\label{fig_tl_rf}
\end{center}
\vskip -0.1in
\end{figure}

\begin{figure}[ht]
\vskip 0.1in
\begin{center}
\centerline{\includegraphics[width=\columnwidth]{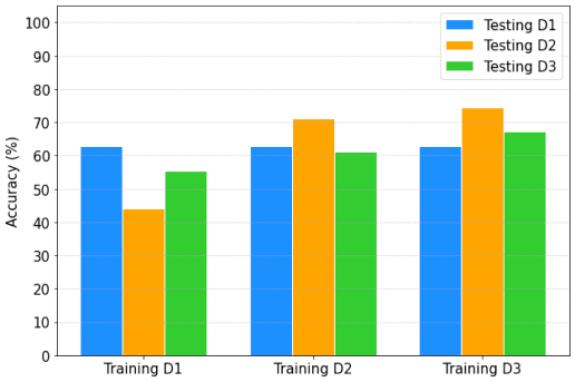}}
\caption{Results with SVM.}
\label{fig_tl_svm}
\end{center}
\vskip -0.1in
\end{figure}

\section{Features involved in the second experiment}
\label{ap-exp2_features}

\begin{table*}[!ht]
\begin{minipage}{\textwidth}
\begin{center}
\caption{Features' comparison.}
\vskip 0.1in
\begin{tabular}{l|ll}
\hline
                                                                           & \multicolumn{1}{l|}{D1}                                                                                                 & D2                                                                                                                                                                                                                     \\ \hline
\multirow{2}{*}{\begin{tabular}[c]{@{}l@{}}Demographic\\ (D)\end{tabular}} & \multicolumn{2}{l}{\begin{tabular}[c]{@{}l@{}}- age: Learner's age\\ - ed\_level: Learner's level of education\end{tabular}}                                                                                                                                                                                                                     \\ \cline{2-3} 
                                                                           & \multicolumn{1}{c|}{/}                                                                                                  & - ed\_field: Learner's field of education                                                                                                                                                                              \\ \hline
\begin{tabular}[c]{@{}l@{}}Academic\\ (A)\end{tabular}                     & \multicolumn{2}{l}{\begin{tabular}[c]{@{}l@{}}- avrg\_grade: Average of the learner's course \\ grades before taking the final exam\end{tabular}}                                                                                                                                                                                                \\ \hline
\begin{tabular}[c]{@{}l@{}}Behavioral\\ (B)\end{tabular}                   & \multicolumn{1}{l|}{\begin{tabular}[c]{@{}l@{}}- sum\_click: Total number of clicks during the \\ courses\end{tabular}} & \begin{tabular}[c]{@{}l@{}}- \%\_completion: Percentage of course completion\\ - nb\_action: Total number of views and achievements \\ during the course\\ - time: Total time learner spent on activities\end{tabular} \\ \hline
\end{tabular}
\vskip -0.1in
\end{center}
\end{minipage}
\end{table*}

\newpage
~
\newpage
\section{Detailed results of the second experiment}
\label{ap-exp2_rez}

\begin{table*}[!h]
\begin{center}
\caption{Results of combinations of usual data sources.}
\label{tab-exp2}
\vskip 0.1in
\begin{tabular}{l|llllllll}
\hline
          & \multicolumn{2}{l}{Accuracy} & \multicolumn{2}{l}{Precision} & \multicolumn{2}{l}{Recall} & \multicolumn{2}{l}{F-score} \\ \cline{2-9} 
          & D1            & D2           & D1            & D2            & D1           & D2          & D1           & D2           \\ \hline
D         & 66.66         & 53.17        & 58.53         & 53.25         & 65.00        & 53.17       & 59.00        & 52.86        \\
A         & 71.66         & 68.56        & 68.33         & 69.21         & 72.50        & 68.56       & 66.49        & 68.29        \\
B         & 70.00         & 69.10        & 72.50         & 69.22         & 72.50        & 69.10       & 70.00        & 69.04        \\
D + A     & 68.33         & 68.85        & 53.33         & 69.57         & 62.50        & 68.85       & 55.16        & 69.56        \\
A + B     & 75.00         & 75.18        & 62.50         & 75.31         & 75.00        & 75.18       & 66.33        & 75.15        \\
B + D     & 68.33         & 69.10        & 54.16         & 69.22         & 65.00        & 69.09       & 57.00        & 69.04        \\
D + A + B & 63.33         & 75.38        & 55.00         & 75.43         & 65.00        & 75.38       & 55.66        & 75.36        \\ \hline
\end{tabular}
\vskip -0.1in
\end{center}
\end{table*}

\newpage
~
\newpage
~
\newpage
\section{Scatter matrix plots}
\label{ap-scatter}

\begin{figure*}[!h]
\vskip 0.1in
     \centering
     \hfill
     \begin{subfigure}
         \centering
         \includegraphics[width=\columnwidth]{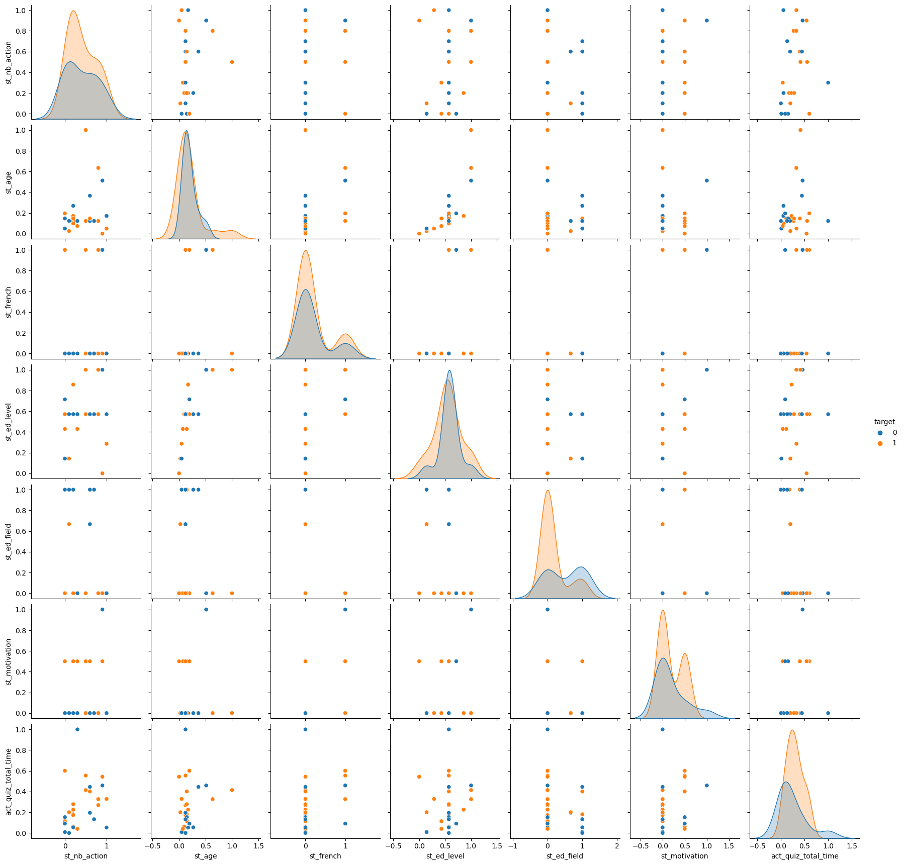}
     \end{subfigure}
     \hfill
     \begin{subfigure}
         \centering
         \includegraphics[width=\columnwidth]{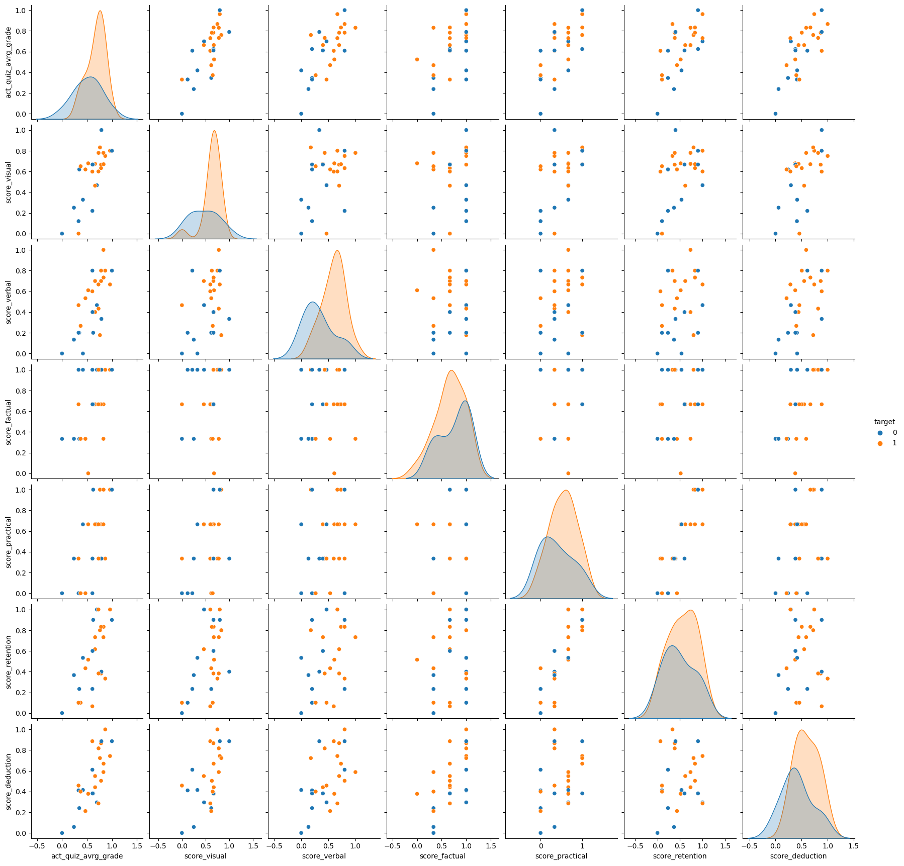}
     \end{subfigure}
        \caption{Scatter matrix plots. (See the close distributions of demographic data according to the target).}
\vskip -0.1in
\end{figure*}

\newpage
\section{Additional sources results}
\label{ap-additional}

\begin{table*}
\begin{center}
\caption{Results of 1-source combinations.}
\label{tab-1source}
\vskip 0.1in
\begin{tabular}{l|llll}
\hline
  & Accuracy (\%) & Precision (\%) & Recall (\%) & F-score (\%) \\ \hline
D & 68.33         & 55.00          & 62.50       & 57.00        \\
A & 76.66         & 72.50          & 77.50       & 72.33        \\
B & 71.66         & 65.00          & 72.50       & 65.66        \\
P & 46.66         & 23.33          & 45.00       & 30.16        \\
L & 61.66         & 54.16          & 65.00       & 54.99        \\ \hline
\end{tabular}
\vskip -0.1in
\end{center}
\end{table*}

\begin{table*}
\begin{center}
\caption{Results of 2-source combinations.}
\label{tab-2source}
\vskip 0.15in
\begin{tabular}{l|llll}
\hline
      & Accuracy (\%) & Precision (\%) & Recall (\%)    & F-score (\%) \\ \hline
D + A & 70.00         & 65.00          & 70.00          & 64.83        \\
D + B & 65.00         & 55.00          & 67.50          & 56.49        \\
D + P & 68.33         & 55.00          & 62.50          & 57.00        \\
D + L & 56.66         & 48.33          & 55.00          & 48.99        \\
A + B & 65.00         & 54.16          & 65.00          & 57.33        \\
A + P & 76.66         & 72.50          & 77.50          & 72.33        \\
A + L & 65.00         & 61.66          & 67.50          & 60.83        \\
B + P & 75.00         & 70.00          & \textbf{77.50} & 70.00        \\
B + L & 66.66         & 57.50          & 67.50          & 58.99        \\
P + L & 58.33         & 49.16          & 60.00          & 50.83        \\ \hline
\end{tabular}
\vskip -0.1in
\end{center}
\end{table*}

\begin{table*}
\begin{center}
\caption{Results of 3-source combinations.}
\label{tab-3source}
\vskip 0.1in
\begin{tabular}{l|llll}
\hline
          & Accuracy (\%) & Precision (\%) & Recall (\%) & F-score (\%) \\ \hline
D + A + B & 60.00         & 51.66          & 60.00       & 53.33        \\
D + A + P & 70.00         & 65.00          & 70.00       & 64.83        \\
D + A + L & 63.33         & 56.66          & 65.00       & 57.50        \\
D + B + P & 65.00         & 55.00          & 67.50       & 56.49        \\
D + B + L & 55.00         & 46.66          & 57.50       & 48.16        \\
D + P + L & 56.66         & 48.33          & 55.00       & 48.99        \\
A + B + P & 63.33         & 52.50          & 62.50       & 54.83        \\
A + B + L & 56.66         & 44.16          & 57.50       & 47.33        \\
A + P + L & 53.33         & 45.83          & 57.50       & 47.49        \\
B + P + L & 60.00         & 49.16          & 62.50       & 51.49        \\ \hline
\end{tabular}
\vskip -0.1in
\end{center}
\end{table*}

\begin{table*}
\begin{center}
\caption{Results of 4-source combinations.}
\label{tab-4source}
\vskip 0.1in
\begin{tabular}{l|llll}
\hline
              & Accuracy (\%) & Precision (\%) & Recall (\%) & F-score (\%) \\ \hline
D + A + B + P & 63.33         & 52.50          & 62.50       & 54.83        \\
D + A + B + L & 51.66         & 41.66          & 52.50       & 43.99        \\
D + A + P + L & 53.33         & 46.66          & 55.00       & 47.49        \\
D + B + P + L & 66.66         & 57.50          & 67.50       & 58.99        \\
A + B + P + L & 56.66         & 44.16          & 57.50       & 47.33        \\ \hline
\end{tabular}
\vskip -0.1in
\end{center}
\end{table*}

\begin{table*}
\begin{center}
\caption{Results of 5-source combinations.}
\label{tab-5source}
\vskip 0.1in
\begin{tabular}{l|llll}
\hline
                  & Accuracy (\%) & Precision (\%) & Recall (\%) & F-score (\%) \\ \hline
D + A + B + P + L & 51.66         & 41.66          & 52.50       & 43.99        \\ \hline
\end{tabular}
\vskip -0.1in
\end{center}
\end{table*}

\begin{figure}[ht]
\vskip 0.1in
\begin{center}
\centerline{\includegraphics[width=\columnwidth]{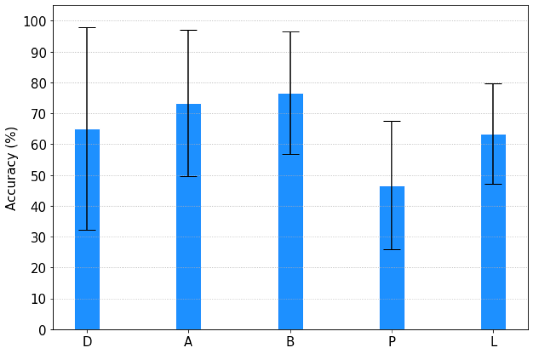}}
\caption{Resulting accuracies with 1-source combinations.}
\label{fig-source1}
\end{center}
\vskip -0.1in
\end{figure}

\begin{figure}[ht]
\vskip 0.1in
\begin{center}
\centerline{\includegraphics[width=\columnwidth]{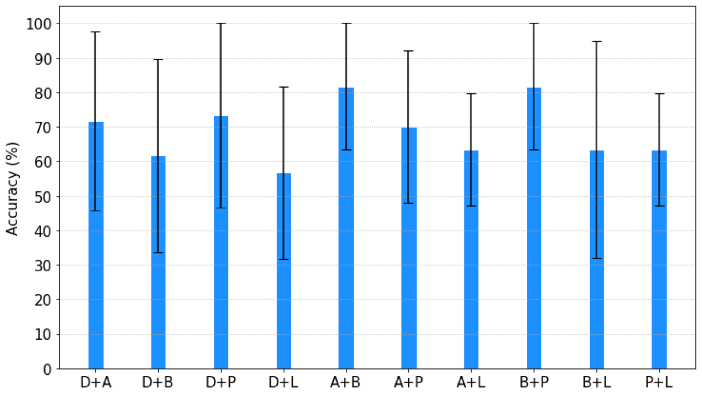}}
\caption{Resulting accuracies with 2-source combinations.}
\label{fig-source2}
\end{center}
\vskip -0.1in
\end{figure}

\begin{figure}[ht]
\vskip 0.1in
\begin{center}
\centerline{\includegraphics[width=\columnwidth]{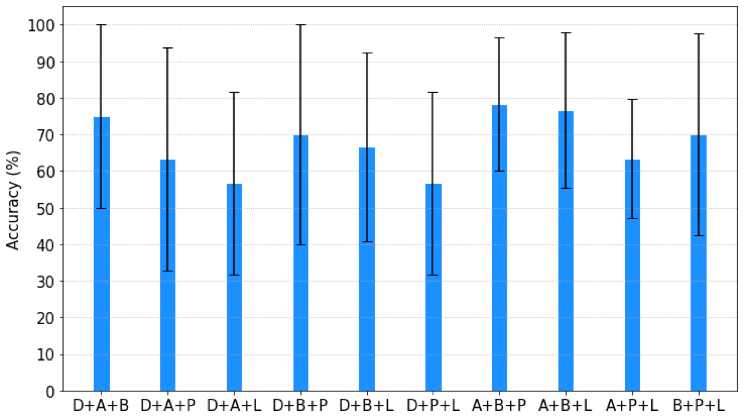}}
\caption{Resulting accuracies with 3-source combinations.}
\label{fig-source3}
\end{center}
\vskip -0.1in
\end{figure}

\begin{figure}[ht]
\vskip 0.1in
\begin{center}
\centerline{\includegraphics[width=\columnwidth]{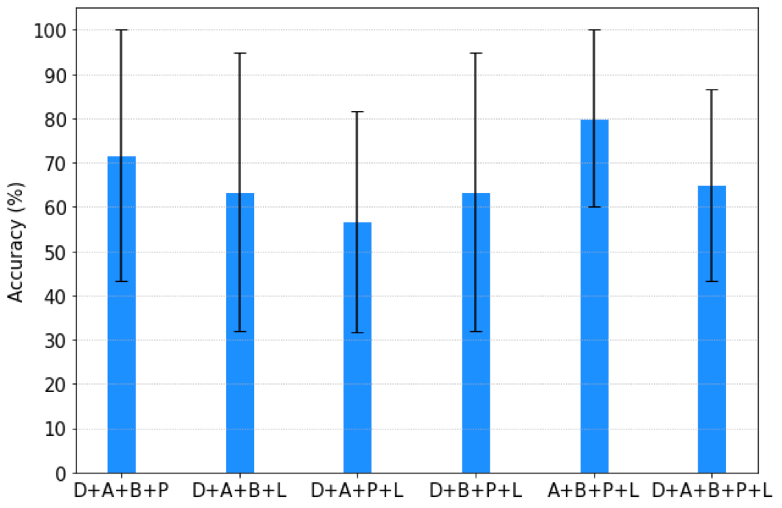}}
\caption{Resulting accuracies with 4- and 5-source combinations.}
\label{fig-source45}
\end{center}
\vskip -0.1in
\end{figure}

\newpage
~
\newpage
~
\newpage
\section{Correlation plots}
\label{ap-corr}

\begin{figure}[ht]
\vskip 0.1in
\begin{center}
\centerline{\includegraphics[width=\columnwidth]{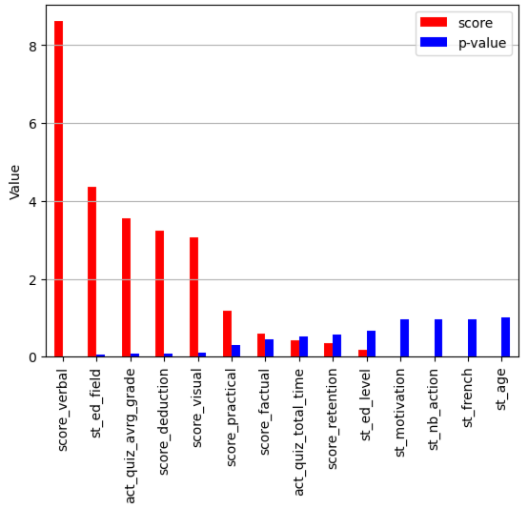}}
\caption{ANOVA F-values.}
\label{fig-anova}
\end{center}
\vskip -0.1in
\end{figure}

\begin{figure}[ht]
\vskip 0.1in
\begin{center}
\centerline{\includegraphics[width=\columnwidth]{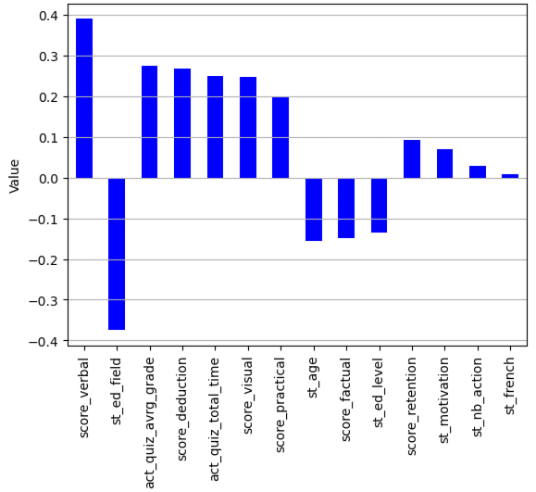}}
\caption{Kendall's correlation coefficients.}
\label{fig-kendall}
\end{center}
\vskip -0.1in
\end{figure}

\begin{figure*}[ht]
\vskip 0.1in
\begin{center}
\centerline{\includegraphics[scale=1]{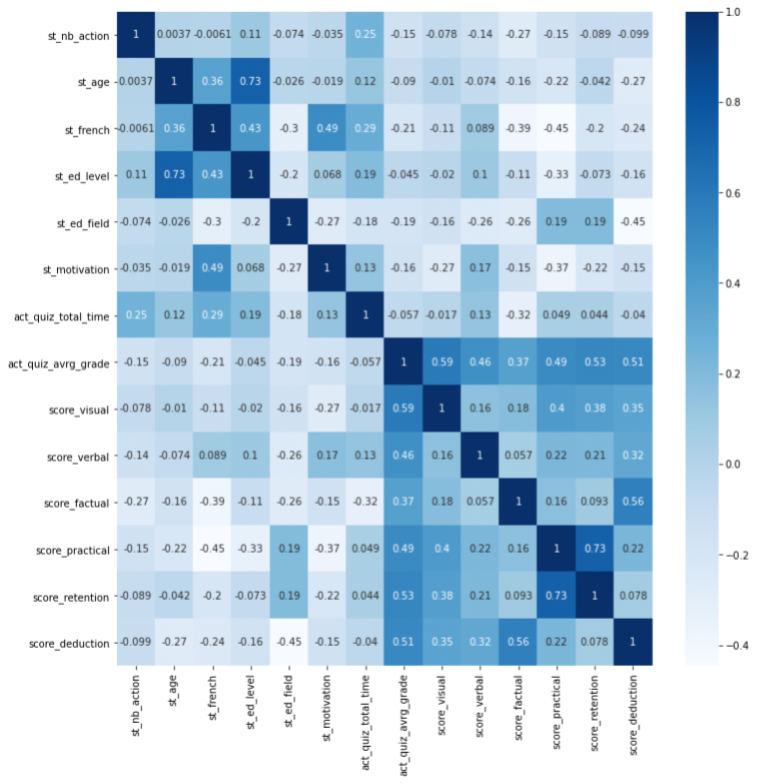}}
\caption{Correlation matrix between features.}
\label{fig-corr_matrix}
\end{center}
\vskip -0.1in
\end{figure*}

%%%%%%%%%%%%%%%%%%%%%%%%%%%%%%%%%%%%%%%%%%%%%%%%%%%%%%%%%%%%%%%%%%%%%%%%%%%%%%%
%%%%%%%%%%%%%%%%%%%%%%%%%%%%%%%%%%%%%%%%%%%%%%%%%%%%%%%%%%%%%%%%%%%%%%%%%%%%%%%

\end{document}